\documentclass[useAMS,usenatbib,usegraphicx]{mn2e}
\usepackage{times}

\title[Hidden starburst in Active Galactic Nuclei]{Hidden starburst in Seyfert 1 galaxies}
\author[A. Rodr\'{\i}guez-Ardila]{A. Rodr\'{\i}guez-Ardila$^{1}$\thanks{Visiting Astronomer at the Infrared Telescope Facility, which is operated by the University of Hawaii under
    contract for the National Aeronautics and Space Administration.} and S. M. Viegas$^{2}$ \\
$^{1}$Laborat\'orio Nacional de Astrof\'{\i}sica - LNA, Rua dos Estados Unidos, 154, CP 21, CEP 37500-000, Itajub\'a, MG, Brazil\\
$^{2}$Instituto de Astronomia, Geof\'{\i}sica e Ci\^encias Atmosf\'ericas - USP, Rua do Mat\~ao 1226, CEP 05508-900, Brazil}
\begin{document}

\date{Accepted: Received ; in original form 2003 January 3}

\pagerange{\pageref{firstpage}--\pageref{lastpage}} \pubyear{2002}

\maketitle

\label{firstpage}

\begin{abstract}
We report the detection of the 3.3\,$\umu$m Polycyclic Aromatic 
Hydrocarbon (PAH) feature in two
Seyfert 1 galaxies $-$ NGC\,3227 and Mrk\,766, and one QSO 
$-$ Mrk\,478, observed with SpeX at IRTF
at a spectral resolution not previously attained for this type of objects.
Except for NGC\,3227, this is the first time that the 
3.3$\mu$m PAH emission is detected in Mrk 766 and Mrk 478.
The widths of the emission, reported also for the 
first time, are rather similar, ranging from 450 \AA\ to 550 \AA.
The luminosity of the 3.3\,$\umu$m PAH emission measured in the
QSO Mrk\,478 places it at a level similar of that found in
starburst and infrared luminous galaxies and implies that
this object is having a vigorous burst of star formation. 
The spatial resolution of the spectra allows us to 
constrain the location of the star-forming region to the inner 
1\,Kpc for the QSO and 150 pc for the Seyferts.
Our results support the idea that 
these objects resides in  molecular gas-rich galaxies and that
their observed  infrared excess is primarily due 
to star formation, as previously indicated by CO and H$_2$ 
observations. We also report, for Mrk\,1239, the presence
of a broad emission feature centred at 3.43\,$\umu$m, 
not previosly detected in an extragalactic object and 
whose origin is not yet clear.
\end{abstract}


\section{Introduction}

Recent investigations of nearby active galactic nuclei
(AGN) have established that a significant fraction of the
energy of Seyfert galaxies may be provided by starburst
activity and that at least some starburst galaxies have
compact AGN cores. As a result, many
observational examples of starburst occurring in active galaxies
have been found \citep{b12, b9}, strengthening
the possible connection between active galaxies and 
circumnuclear starburst activity.

Among the many indicators that may signal the presence of a
circumnuclear starburst in AGN are stellar absorption 
features in ultraviolet spectra
(e.g., \citealt{b6, b5}),
emission signatures of Wolf-Rayet stars \citep{b10};
Ca\,{\sc{ii}} triplet absorption at 
8555 \AA\ \citep{b26, b15} and the presence of a featureless continuum 
\citep{b3}. However, except for a very few
cases, circumnuclear starburst has rarely been detected  
in Seyfert 1 galaxies using the above techniques. Most 
of the detections came from Seyfert 2 galaxies. 

A much more sensitive tracer of powerful starburst
activity, proposed by \citet{b17}, is the mass to light
ratio at 1.65\,$\umu$m ($M/L_{H}$), derived from the observed
stellar flux and velocity dispersion  of the stellar absorption
features of Si at 1.59\,$\umu$m, CO(6,3) at 1.62\,$\umu$m, and
CO(2,0) at 2.29\,$\umu$m. Although this test
has been successfully applied to obscured AGN (i.e$-$ Seyfert 2's),
it has lead, again, to the non-detection of starburst activity in 
genuine Seyfert 1 galaxies. In the light of the unified model for AGN 
\citep{b1}, the above results are not readily explained. 
If there exist a connection between AGN and starburst, one 
would expect to observe an even distribution of circumnuclear
star-forming regions among the two types of Seyferts. Is the 
lack of a starburst
component in type 1 AGN a real effect or it does exists but
is hidden by the much stronger radiation from the central
engine?

In the last decade, spectroscopy in the wavelength range 3$-$4\,$\umu$m 
has proben to be an effective way to detect the presence of 
starburst activity by means of the emission features at 
3.3\,$\umu$m and sometimes at
3.4\,$\umu$m \citep{b27, mi, b8, b7}. The advantage of 
observing in this region is clear. In the UV and optical bands, 
emission from AGN in Seyfert 2s is much weaker than in Seyfert 
1s due to large dust extinction, so detection of compact 
starbursts is technically much easier in the former than in 
the later. However, at 3--4\,$\umu$m, dust extinction is much 
lower (A$_{3--4\umu m}$/A$_{\rm v}$ = 0.06; \citealt{rl85}). 
Flux attenuation of AGN emission at 3--4\,$\umu$m differs only 
by a factor of 2-5 between Seyfert 1s and Seyfert 2s if
A$_{\rm v}$ toward the obscured AGN in the later is 10-30 mag 
(e.g., \citealt{fa98}). Thus, at 3--4\,$\umu$m, detection of 
the compact nuclear starbursts in Sy1s is not so difficult 
compared to Sy2s. Given that the compact nuclear starbursts 
have been detected in Sy2s, their detection in Sy1s is also 
feasible.  

Emission features at 3.3\,$\umu$m and 3.4\,$\umu$m has been 
attributed to large molecules like polycyclic 
aromatic hydrocarbon (PAH; \citealt{b11}) illuminated by
ultraviolet photons from early type stars.  They have
been found to be prominent in pure starburst, infrared 
luminous galaxies and obscured AGN \citep{b13, b8, b7}
and are associated 
to star formation activity in galaxies. The lack of 
available spectroscopic observations in Seyfert 1 prevents 
to state if this test also fails in these objects. To 
our knowledge, Mrk\,231 has been the only Seyfert 1 galaxy
observed in the 3$-$4\,$\umu$m region with the goal of 
detecting signals of starburst activity. A positive
detection of the 3.3\,$\umu$m PAH feature in that object
\citep[][hereafter ID00]{b8} makes NIR spectroscopy a 
promising technique. Nonetheless, it should be keep in mind that 
in the close universe, Mrk\,231 is the brightest active 
galaxy in the IR ($L_{\rm IR}$=3.5 $\times$
10$^{12}$ L\sun). Moreover, the 
existence of a strong OH maser source in its nucleus 
\citep{b2} points, in fact, towards the existence of a strong 
starburst in its circumnuclear region.  

This letter reports the results of 3$-$4\,$\umu$m 
spectroscopy aimed at detecting hidden starburst 
activity in type 1 AGN and quantifying its importance
to the IR energetics. In \S ~\ref{obs}
we describe the observations, data reduction and
main results. A discussion of our data is in \S~\ref{disc}.
Additional evidence to support our main findings is 
in \S~\ref{evid}. Conclusions are in \S~\ref{final}.
Throughout this paper, a Hubble constant of H$_{\rm o}$=
75 km\,s$^{-1}$\,Mpc$^{-1}$ will be used.

\section{Observations, Data Reduction and Results} \label{obs}

\subsection{Observations}

Five Seyfert 1 galaxies and one QSO\footnote{The
terms ``Seyfert 1'' and ``QSO'' are  used
throughout the text to indicate differences in the 
absolute magnitude of the AGN. Objects brighter than 
M$_{\rm B}$=-23 are classified as QSOS and those
fainter than that value are classified as Seyfert 1s. Spectroscopically, 
both types of objects are indistinguishable from each other.} were observed
during the night of April 22, 2002 using the SpeX facility
spectrometer \citep{b19} at the 3.0 m NASA infrared 
telescope facility  on Mauna Kea, Hawaii. In the UV and 
optical region, the spectra of the sources
is dominated by the non-stellar continuum radiation
emitted by the AGN as well as emission features emitted by 
the broad line region. Except for NGC\,3227, no significant  
contribution of circumnuclear stellar population
has been reported. Table~\ref{basic} provides useful information
for the galaxies.

In the long wavelength
cross-dispersed mode, SpeX provides nearly continuous coverage
from 2.2\,$\umu$m through 4.2\,$\umu$m. Here, we will discuss
only the spectra covering the 3.0$-$3.6\,$\umu$m region. They
were acquired using a 0$\arcsec$.8 slit width, oriented north-south
yielding an instrumental resolution R$\sim$ 1000. Observations 
were done nodding in an off-on-on-off source 
pattern with a typical individual integration time of 30 seconds
(usually 10 coadds of 3 sec each) and
total integration times (excluding overheads) between
30 minutes and 50 minutes.  A0V stars were observed
near targets to provide telluric standards at similar airmasses.
Seeing, as measured in the K band, varied between 
0$\arcsec$.7$-$0$\arcsec$.8 (FWHM) during the night.

\begin{table*}
\centering
 \begin{minipage}{140mm}
  \caption{Useful information, $IRAS$ Fluxes and infrared luminosity for the galaxy sample.} 
 \label{basic}
  \begin{tabular}{@{}lcccccccc@{}}
  \hline
    &  & R   & $f_{12}$ & $f_{25}$ & $f_{60}$ & $f_{100}$ & log$L_{\rm IR}$\footnote{$L_{IR}$=$L(8-1000~\umu m)$ is computed using the flux in all four $IRAS$ bands according to the
    prescription given by \citet{b21}.} & log$L_{\rm FIR}$\footnote{Logarithm of the far-infrared luminosity (40-500\,$\umu$m) computed using the flux in the 60\,$\umu$m and 100\,$\umu$m $IRAS$ bands according to the
    prescription given by \citet{b21}.}  \\
Object & $z$ & (pc) & (Jy) & (Jy) & (Jy) & (Jy) & (erg s$^{-1}$) & (erg s$^{-1}$) \\
  (1) & (2) & (3) & (4) & (5) & (6) & (7) & (8) & (9) \\
\hline
1H\,1934-063  & 0.01059 & 143 & 0.50$\pm$0.04  & 1.06$\pm$0.03 & 2.81$\pm$0.14 &  4.75$\pm$0.48 & 43.96 &  43.80 \\
MRK\,478  & 0.07906 & 1140 & 0.12$\pm$0.003 & 0.19$\pm$0.02 & 0.57$\pm$0.04 &  0.92$\pm$0.11 & 44.96 &  44.79 \\
MRK\,766  & 0.01293 & 150 & 0.39$\pm$0.04  & 1.30$\pm$0.03 & 4.03$\pm$0.28 &  4.66$\pm$0.28 & 44.18 &  44.07 \\
MRK\,1239 & 0.01993 & 308 & 0.65$\pm$0.07  & 1.14$\pm$0.05 & 1.33$\pm$0.11 &  $<$2.41 & 44.44 &  44.03 \\
NGC\,3227 & 0.00386 & 60 & 0.67$\pm$0.05  & 1.76$\pm$0.04 & 7.83$\pm$0.39 &  18.4$\pm$0.15 & 43.49 &  43.46 \\
NGC\,4051 & 0.00234 & 36 & 0.86$\pm$0.05  & 1.59$\pm$0.05 & 10.4$\pm$0.05 &  23.9$\pm$1.20 & 43.11 &  43.11 \\
\hline
\end{tabular}
\end{minipage}
\end{table*}

\subsection{Data Reduction}

 The spectral extraction and calibration procedures were
 carried out using Spextool, the in-house software developed
 and provided by the SpeX team for the IRTF
 community\footnote{Spextool is available from the
 IRTF website: http://irtf.ifa.hawaii.edu/Facility/spex/spex.html}.
 Each on-off source pair of observations was reduced individually
 and the results summed to provide a final spectrum.  A 1$''$.6 
$-$ 1$''$.8 
 diameter aperture was used to extract the spectra. This window 
 size integrates nearly all counts detected in the light profile, 
 which was essentially point-like and indistinguishable from that of
the A0V standards.  Observations of an argon arc lamp enabled 
 wavelength calibration
 of the data.

 Wavelength calibrated final target spectra were divided
 by AOV stars observed at a similar airmasses. The spectra were 
 each multiplied by a black body function corresponding to 
 10,000 K in order to restore the true continuum shape of the 
 targets. Spectra were flux calibrated by normalizing to the $L$ band
 magnitude of the corresponding A0V standard star observed. 
 Finally, the spectra were corrected for Galactic extinction, as 
 determined from the {\it COBE/IRAS} infrared maps of 
 \citet{b24} and then 
 shifted to rest wavelength using the $z$ values available in NED.

\subsection{Results}

\begin{figure}
\includegraphics[width=8.5cm]{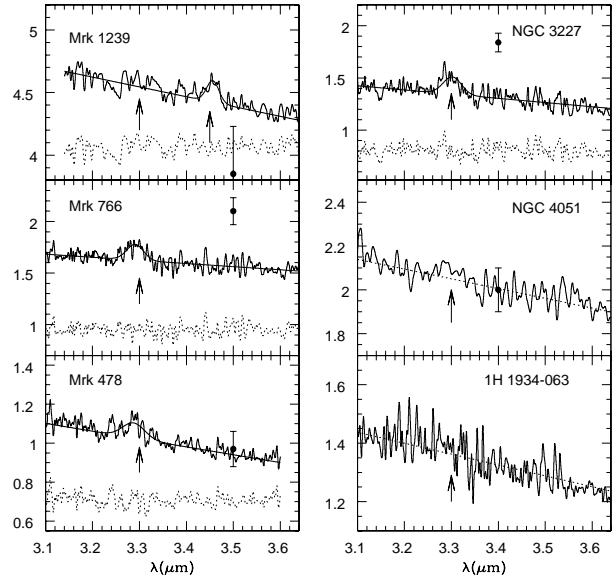}
\caption{Flux calibrated spectra of the galaxy sample. 
The abscissa is in rest wavelength ($\umu$m) and the 
ordinate the flux in units of 10$^{-15}$ erg cm$^{-2}$ 
s$^{-1}$ \AA $^{-1}$. The arrow indicates the centre of 
the 3.3\,$\umu$m PAH feature as determined from the 
Gaussian fit or the expected position in the objects 
where only upper limits were derived. The second arrow
at 3.45\,$\umu$m in Mrk\,1239 marks the position of
the 3.43\,$\umu$m emission band (see text for further details).
The dashed line is the residual after subtraction of 
the continuum plus Gaussian fit. It was arbitrarily 
displaced in the y-axis for displaying purposes. 
The error bar is the $L$ photometry taken from \citet{b9} 
for NGC\,3227, NGC\,4051 and Mrk\,766 (aperture of 
8.6$\arcsec$ for the former two and 5.4$\arcsec$ for 
the later object); \citet{b16} for Mrk\,478 (aperture 
of 5.5$\arcsec$); and \citet{b20} for Mrk\,1239 (aperture 
of 8.6$\arcsec$). Note that for this later object, L 
photometry was derived from K photometry.}
\label{fig1}
\end{figure}

The resulting flux-calibrated spectra are 
displayed in Figure~\ref{fig1}. In order to increase the
S/N, data for Mrk\,1239 and NGC\,4051 were smoothed using
a Gaussian filter of 5\,\AA\ FWHM. The 3.3\,$\umu$m PAH feature is 
clearly present in the QSO Mrk\,478 and in the two 
Seyfert 1 galaxies Mrk\,766 and NGC\,3227. In the
remaining three objects, evidence of that emission is uncertain,
except probably in NGC\,4051, where a feature at the 
expected position is seen.
For Mrk\,478 and Mrk\,766 it is the first time that 
the 3.3\,$\umu$m  PAH emission is detected. 
Imanishi (2002, hereafter IM02) had already reported the 
detection of the 3.3\,$\umu$m emission in NGC\,3227.
Nonetheless, they considered this AGN as Seyfert 2
while spectroscopically it is classified as
Seyfert 1 \citep{b23}. We keep the latter classification.
and confirm the presence of the 3.3\,$\umu$m PAH
emission at a higher spectral and spatial resolution. 

Table~\ref{pah_flux} lists the integrated 3.3\,$\umu$m PAH 
fluxes and the corresponding equivalent widths and luminosities. 
Uncertainties are 3$\sigma$ representive. Fluxes were measured
in two ways: integrating directly over the feature and fitting
a Gaussian profile. Both methods gave similar results within
the uncertainties set by the S/N. The continuum
level was determined by a linear fit of the data points that
are free of emission or absorption over the interval 
3.0$-$3.7\,$\umu$m. The peak of the PAH feature as
determined from the Gaussian fit was 3.29\,$\umu$m
for Mrk\,478 and 3.30\,$\umu$m for NGC\,3227 and Mrk\,766,
in agreement with the values reported in the
literature for starburst and Seyfert 2 galaxies (\citealt{mi};
ID00; IM02). The FWHM of the 3.3\,$\umu$m PAH emission are rather 
similar: 550\,\AA, 460\,\AA\ and 450\,\AA\ for Mrk\,478, Mrk\,766 and
NGC\,3227, respectively. They agree with the value of 
420\,\AA\ reported by \citet{b27} for extended sources 
such as planetary nebulae and H\,{\sc ii} regions.
No previous measurements of the 3.3\,$\umu$m PAH width for
Seyferts galaxies have been reported in the literature so a 
direct comparison with our data is not possible. However, a visual
inspection of our spectra and those of ID00 and IM02 allows
us to conclude that the FHWM of the 3.3\,$\umu$m feature are
rather similar. 

\begin{table}
\centering
 \begin{minipage}{140mm}
  \caption{Measured properties of the 3.3\,$\umu$m PAH feature.} 
 \label{pah_flux}
  \begin{tabular}{@{}lcccccc@{}}
  \hline
Object & $f_{3.3}$\footnote{in units of 10$^{-14}$ erg s$^{-1}$ cm$^{-2}$} & EW$_{3.3}$ & $L_{3.3}$\footnote{in units of erg s$^{-1}$}  & $L_{3.3}/L_{\rm IR}$ \\
        &    & (\AA) & ($\times$ 10$^{39}$) &   ($\times10^3$) \\
(1)  & (2) & (3) & (4) & (5) \\
\hline
1H1934  & $<$1.4       & $<$10     &  $<$3.04      & $<$3.3 $\times 10^{-2}$ \\
MRK478  & 4.3$\pm$0.8  & 41$\pm$8  &  512$\pm$83   & 5.6$\pm$0.9$\times 10^{-1}$ \\
MRK766  & 6.5$\pm$1.5  & 40$\pm$11 &  20.7$\pm$4.8 & 1.4$\pm$0.4$\times 10^{-1}$ \\
MRK1239 & $<$1.2       & $<$2.6    &  $<$8.9       & $<$3.2$\times 10^{-2}$    \\
NGC3227 & 7.0$\pm$1.20 & 52$\pm$10 &  2.1$\pm$0.4  & 6.8$\pm$1.3$\times 10^{-2}$ \\
NGC4051 & $<$2.9       & $<$14     &  $<$0.30      & $<$2.3$\times 10^{-2}$    \\
\hline
\end{tabular}
\end{minipage}
\end{table}

The small aperture of 0.8\arcsec $\times$ 0.8\arcsec\ employed
for integrating the spectra sets tightly constraints to the
distance from the centre of the AGN to the starburst. 
For each object of our sample, an upper limit to this distance 
is listed in column 3 of Table~\ref{basic}. In NGC\,3227 it
corresponds to the inner 60 pc; in Mrk\,766, the inner 150 pc and in
Mrk\,478, the inner 1\,kpc. These values represents,
by far, the best observational estimates on the location 
of a circumnuclear starburst in Seyfert 1 galaxies and
QSOs.

In Mrk\,1239, NGC\,4051 and 1H\,1934-063, no clear 
evidence for the 3.3\,$\umu$m PAH emission 
is observed and only upper limits to its flux could be derived 
(see Table~\ref{pah_flux}). The lack
of PAH emission is interpreted as either they are unobscured 
AGN with no significant circumnuclear star-formation 
activity or if there is 
any starburst component, it is located outside the inner 
few hundred parsecs of the nucleus. A third possibility 
may be related to the correlation between AGN 
activity and active star-forming host proposed by 
IM02: the most powerful 
monsters live in the more actively star-forming host 
galaxies. For instance, it is well known that NGC\,4051 
is one of the weakest Seyfert 1 nuclei \citep{b5}. 
If the correlation is valid, 
it would imply a very low level of star formation in this
object, probably outside the detection limits. The same 
argument can be applied to 1H\,1934-063, another low  
luminosity Seyfert 1 galaxy.

It is important to note the presence of an emission
feature at 3.45$\umu$m in the spectrum of Mrk\,1239,
as can be seen in the upper left panel of Figure~\ref{fig1}. 
In the remaining 5 objects, none is seen at this 
position, either in emission or in absorption. No previous 
report of such an emission were found in the literature
for an extragalactic object. Reports of an absoption
feature at 3.4$\umu$m, attributted to carbonaceous dust grain,
exist for Seyfert 2 (\citealt{im00a, im00b}; ID00;
IM02) and composite starburst/obscured AGN galaxies \citep{im01}. 
In addition, \citet{b27} detected an emission
feature of unclear origin at 3.43\,$\umu$m in the spectrum 
of Elias\,1, a Herbig Ae source. Whether the emission seen
in Mrk\,1239 is related to the 3.4$\umu$m absorption or
to the stellar 3.43\,$\umu$m emission is not possible to
tell from our data. Nonetheless, the fact that no clear 
evidence of the 3.3\,$\umu$m PAH emission is seen on this
source and that it is the only object in which the 
3.4\,$\umu$m is observed gives support to the hypothesis 
that the 3.3\,$\umu$m and 3.4\,$\umu$m features are 
unrelated \citep{b27}.
\section{Discussion} \label{disc}

Several important points arise from the data listed in
Tables~\ref{basic} and~\ref{pah_flux}, mainly when compared 
to the information available in the literature for pure 
starburst and Seyfert 2 galaxies. 
 
Even in the presence of starburst activity, small 3.3\,$\umu$m
PAH equivalent widths (hereafter EW$_{3.3}$) 
in AGN is most likely caused by the strong
continuum emitted by circumnuclear dust heated by the 
central source, which peaks at the 3\,$\umu$m to 4\,$\umu$m
region. For that reason, it is not surprising to find that the
average EW$_{3.3}$ measured in our galaxy sample is 44$\pm$10 \AA,
excluding the upper limits detections, and 27 \AA,
including them. Considering that starburst-dominated 
galaxies show EW$_{3.3}$ in the interval 600$-$2400 \AA\ 
(ID00), we deduce that more than 
90\% of the continuum flux in the 3.3\,$\umu$m
region originates from dust heated by the AGN in Seyfert 1. 
For Seyfert 2 galaxies, nearly 80 \% of the 3--4 $\umu$m 
continuum is also related to AGN activity (IM02). 
However, low EW$_{3.3}$ does not necessarily means that the 
IR emission (8$-$1000\,$\umu$m) is also 
dominated by the AGN contribution.  

A first approach to estimate the relative importance of a
hidden starburst is to compare the IR and 3.3\,$\umu$m 
luminosities. Following \citet{b21},
the L$_{IR}$ is obtained using  the 12\,$\umu$m, 25\,$\umu$m, 
60\,$\umu$m and 100\,$\umu$m $IRAS$ fluxes, also listed
in Table~\ref{basic}. The L$_{3.3}$/L$_{IR}$ ratios derived 
for the objects of our sample  are shown in column 5 of 
Table~\ref{pah_flux}. We found 
that they are rather similar to those measured in the
Seyfert 2 galaxy sample of IM02.
The most striking result, however, is for Mrk\,478. 
The L$_{3.3}$/L$_{\rm IR}$ ratio for this source 
reaches 0.56$\times$10$^{-3}$, more than half of the
typical value found for starburst dominated galaxies 
(L$_{3.3}$/L$_{\rm IR} \sim$1$\times$10$^{-3}$; \citealt{mo90}).
The assumption that 0.1\% of the IR luminosity in starburst
galaxies is converted into 3.3$\umu$m emission 
has later been supported by several works (i.e., \citealt{mi}; IM02).
The above means that Mrk\,478 harbors a powerful hidden
compact starburst region, not detected by optical and/or
UV spectroscopy. 

Mrk\,766 and NGC\,3227, the other two Seyfert 1 galaxies
with clear identification of the 3.3\,$\umu$m feature
show L$_{3.3}$/L$_{\rm IR}$ ratios typical of those measured
in Seyfert 2 galaxies. As for Mrk\,478, Mrk\,766 has no signs
of circumnuclear star formation in the visible or UV region.
This suggests that the lack
of detection of starburst activity in Seyfert\,1 galaxies
and QSOs by means of UV and optical indicators can be 
due to the strong contribution of the AGN 
radiation to the observed continuum rather than to a truly 
deficit of star-forming regions in these objects. 
A more definite conclusion on this issue should be reached 
as soon as observations of a larger number of objects 
become available.  But it is
clear that  had we applied any of
the diagnostics usually employed to search for starburst in
AGN (see \S 1) only one out of the six galaxies of our sample
(NGC\,3227) would present signs of active star formation. 

\section{Molecular CO and the PAH emission: 
further evidences for starburst in Type 1 objects} \label{evid}
 
Several pieces of evidence give additional support to 
our results and suggest a tight correlation between
the 3.3\,$\umu$m PAH and molecular CO emission. 
\citet{b4} observed a sample of 10 
infrared-excess optically selected QSOs from the 
Palomar-Green (PG) Bright Quasar Survey with the goal 
of establishing whether or not quasi-stellar objects reside
in molecular-rich galaxies as well as determining how the
infrared and molecular gas properties of QSOs compare with
those of ultraluminous infrared galaxies. Mrk\,478 is among
the objects showing millimetre CO\,(1$\rightarrow$0) emission. 
According to \citet{b4} results, this QSO shows one of 
the largest CO luminosities and a very high H$_{2}$
molecular mass (8.0 $\times$ 10$^{9}$ M\sun). It
means that Mrk\,478 lives in a host galaxy containing
significant supply of molecular gas, capable of igniting
and sustaining both AGN and star formation activity.
All the QSOs of \citet{b4} sample have high 
values of $L_{\rm IR}/L_{\rm CO}$. In starburst
galaxies, that ratio is commonly referred to as the star 
formation efficiency. Large values of $L_{\rm IR}/L_{\rm CO}$ 
in AGN can be interpreted as due to either dust heating by the 
QSO or to star formation intimately connected to the QSO
phenomenon. The presence of the 3.3\,$\umu$m feature in 
Mrk\,478 and the high L$_{3.3}$/L$_{\rm IR}$ ratio,  
similar to that observed in 
IRLGs and pure starburst objects, clearly points out to the 
second possibility. In addition, if this hypothesis is 
correct and the CO emission comes primarily for the starburst
component, our data sets important constraints to the
region where it is located: the inner 1\,kpc.

Similar results are obtained for NGC\,3227. 
Using NICMOS imaging, \citet{b18} 
found an extended H$_{2}$ emission that lies in a 100\,pc 
diameter ``disk'' with a major P.A $\sim$100$^{o}$,
probably associated with large quantities of dense 
molecular material. Variations in the distribution and
velocity components observed in [O\,{\sc{iii}}] and 
H$\alpha$+[N\,{\sc{ii}}] on this source by \citet{b5}
suggest that star formation is vigorous in its central 
region. In addition, from the CO velocity field maps,
\citet{b23} found evidence that a 
warped molecular disk is present in the inner 70\,pc 
of this galaxy. Our 
3.3\,$\umu$m PAH detection in NGC\,3227 is in total 
agreement with the above results and confirms the
presence of a circumnuclear starburst in this galaxy.
Moreover, the size of the emitting PAH region, 
constrained to the central 60\,pc, is coincident with 
the size of the CO molecular disk, suggesting a close 
connection between these two features.  

CO observations of Mrk\,766 have failed to detect 
significant emission of molecular gas in this object. 
However, \citet{b25}  derived an upper limit
for its H$_{2}$ mass and found it to be 
4.4 $\times$ 10$^{8}$ M\sun, almost 20 times lower
than that of Mrk\,478. Nonetheless, CO observations of 
NGC\,3227 by these same authors allowed them to 
derived an H$_{2}$ mass of 1.6 $\times$ 10$^{8}$ M\sun,
3 times smaller than the upper limit found for
Mrk\,766. Putting together the CO and 3.3\,$\umu$m 
PAH measurements for Mrk\,478, Mrk\,766 and NGC\,3227
it appears that they are intimately related and
connected to the star formation phenomenon. 

\section{Summary} \label{final}

The apparent lack of circumnuclear starburst in type 1 AGN
has been analysed using medium resolution spectroscopy 
in the near-infrared for 5 Seyfert 1 galaxies and one QSO. 
A clear identification of the 3.3 $\mu$m PAH emission, 
usually associated to star formation activity, was obtained 
in  Mrk 478, Mrk\,766 and NGC\,3227. Upper limits for the 
3.3\,$\umu$m flux was derived for the remaining 
objects. The values of the 3.3\,$\umu$m PAH 
equivalent widths are very similar to those found 
in Seyfert\,2 galaxies known to host a 
circumnuclear starburst. It suggests that AGN of 
type 1 and 2 present a comparable level of 
star-formation activity but probably
in the former type of objects it is hidden by the 
direct view of the central radiation.
 
Although more than 90\% of the continuum flux at 
3.3\,$\umu$m was found to be associated to the AGN, 
in Mrk\,478 and Mrk\,766 the L$_{3.3}$/L$_{\rm IR}$ 
ratio, a measurement of  the energetic contribution 
of the starburst to the 8$-$1000\,$\umu$m continuum, is
comparable to the values measured in infrared
luminous galaxies and pure starburst galaxies. This
result indicates that the IR luminosity in these
two objects is significantly powered by a starburst
component. CO observations in the millimetre
region support this scenario. A large mass of
molecular gas is derived in the objects with clear
identification of the 3.3\,$\umu$m PAH feature.
This gas is most probably located in the same region 
where the PAH is emitted, providing significant fuel 
capable of igniting and sustaining both AGN and 
star formation activity. 

\section*{Acknowledgments}
This research have been supported by the Funda\c c\~ao de
Amparo \`a Pesquisa do Estado de S\~ao Paulo (FAPESP) to
S.M.V (proc. 00/06695) and PRONEX grant 662175/1996-4 to 
S.M.V and A.R.A. This research has made use of the NASA/IPAC
Extragalactic Database (NED) which is operated by the Jet
Propulsion Laboratory, California Institute of Technology,
under contract with the National Aeronautics and Space
Administration. We thank Dr. Masatoshi Imanishi, the referee,
for useful suggestions which helped to improved this letter.

\end{document}